\documentstyle[psfig]{l-aa}
\input{epsf}

\begin{document}

    \thesaurus{ 10.08.1: 13.07.2: 12.04.1 }

\def\figinsert#1#2{\epsfbox{#1} \message{#2} }          %insert figures 
\def\la{\lower.5ex\hbox{$\; \buildrel < \over \sim \;$}}
\def\ga{\lower.5ex\hbox{$\; \buildrel > \over \sim \;$}}
\def\r{\hangindent=1pc \noindent}
\def \ang{{\rm \AA}}
\def \hi {H\,{\sc i~}}
\def\h{H~I}
\def\he{He~I}
\def\hei{He \,II}
\def\12{{1\over 2}}
\def\msun{\rm {M_{\odot}}}
\def\vsun{v_{\odot}}
\def\lsun{L_{\odot}}
\def\div{\nabla\cdot}
\def\grad{\nabla}
\def\rot{\nabla\times}
\def\eg{{\it e.g.,~}}
\def\ie{{\it i.e.,~}}
\def\etal{{\it et~al.,~}}
\def\de{\partial}
\def\ltsima{$\; \buildrel < \over \sim \;$}
\def\simlt{\lower.5ex\hbox{\ltsima}}
\def\gtsima{$\; \buildrel > \over \sim \;$}
\def\simgt{\lower.5ex\hbox{\gtsima}}
\def\etal{{et~al.}}
\def\noi{\noindent}
\def\bs{\bigskip}
\def\ms{\medskip}
\def\ss{\smallskip}
\def\ob{\obeylines}
\def\l{\line}
\def\hrf{\hrulefill}
\def\hf{\hfil}
\def\q{\quad}
\def\qq{\qquad}
\renewcommand{\deg}{$^{\circ}$}
\newcommand{\um}{$\mu$m}
\newcommand{\uk}{$\mu$K}
\newcommand{\qrms}{$Q_{rms-PS}$}
\newcommand{\n}{$n$}
\newcommand{\cdmr}{${\bf c}_{\rm DMR}$}
\newcommand{\xrms}{$\otimes_{RMS}$}
\newcommand{\gt}{$>$}
\newcommand{\lt}{$<$}
\newcommand{\ldl}{$< \delta <$}
\newcommand{\be}{\begin{equation}}
\newcommand{\ee}{\end{equation}}
\newcommand{\ba}{\begin{eqnarray}}
\newcommand{\ea}{\end{eqnarray}}

%%%%%%%%%%%%%%%%%%%%%%%%%%%%%%%%%%%%%%%%%%%%%%%%%%%%%%%%

%%%%%%%%%%%%%%%%%%%%%%%%%%%%%%%%%%%%%%%%%%%%%%%%%%%%%%%%
\title{H$_2$ dark matter in the galactic halo from EGRET}
 
\author{P. M. W. Kalberla$^1$, Yu. A. Shchekinov$^{2,3}$,  R.-J. 
Dettmar$^4$ }
\bs
\institute
{$^1$Radioastronomisches Institut, Universit\"at Bonn, D-53121 Bonn, 
Germany\\
$^2$Department of Physics, Rostov State University,
344090 Rostov on Don, Russia\\
$^3$Osservatorio Astrofisico di Arcetri, I-50125 Florence, Italy\\
$^4$Astronomisches Institut, Ruhr-Universit\"at Bochum, 
D-44780 Bochum, Germany
}
\bs
%\email{( yus@rsuss1.rnd.runnet.su, yuris@arcetri.astro.it)}

     \date{Received 8 July 1999; accepted 1 September 1999}

     \maketitle
%\maintitlerunninghead{H$_2$ dark matter in the galactic halo from EGRET }
%\authorrunninghead{{P. M. W. Kalberla, Yu. A. Shchekinov, and R.-J. Dettmar }
\markboth{P.M.W. Kalberla, Yu.A. Shchekinov, and R.-J. Dettmar: 
H$_2$ dark matter in the galactic halo from EGRET }{}

% ------------------------------ Title page ---------------------------------
 
\begin{abstract}

We present a model for the interpretation of the $\gamma$-ray 
background emission from the Galactic halo, 
which we attribute to interaction of high-energy 
cosmic rays with dense molecular clumps. In a wide range of clump 
parameters we calculate the expected $\gamma$-photon flux, 
taking into account for the first time optical depth effects. 
This allows us to derive new constraints 
on masses and sizes of possible molecular clumps and 
their contribution to the galactic 
surface density. 

The observed $\gamma$-ray distribution can be explained best by 
models with a flattened halo distribution of axis ratio $\sim 0.3$. 
If optical depth effects are important, the clumps must have radii of 
$\sim 6$ AU and masses of $\sim 10^{-3} \msun$. This would result in a 
total mass of $\sim 2 \cdot 10^{11} \msun$ for such clumps and contribute 
with $\Sigma \sim 140 \; {\rm M_{\odot}\; pc^{-2}}$ to the 
local surface density.

\end{abstract}


%-------------------------- Section I ----------------------------

\section{Introduction}

Recent observational results of the anisotropy of the cosmic microwave 
background, deuterium abundance from cosmological nucleosynthesis, 
dynamics of clusters of galaxies and the data from the Supernova 
Cosmology Project provide evidence for the distribution of mass in the 
universe in the proportion: $\Omega_{\rm b}\simeq 0.05$ for baryons, 
$\Omega_{\rm m}\simeq 0.35$ for non-baryonic matter, and $\Omega_\Lambda\simeq 0.6$ 
for the cosmological constant (Turner, 1999). Since luminous baryons contain 
only $\Omega{\rm _b^L}\simeq 0.007$, more than 85 \% of the baryonic mass is still 
undetected. Pfenniger \& Combes (1994) and more recently Walker \& Wardle (1998) 
have suggested that a considerable fraction of baryonic mass can be contained in 
dense molecular clumps of AU sizes and about a Jovian mass, which in turn may 
manifest themselves through extreme scattering events (ESEs) -- dramatic flux 
changes of compact radio quasars over several weeks (Fiedler et al., 1987). 
However, estimates of this fraction are uncertain because the existing 
statistics of the ESEs sets a wide range for the covering factor of the 
radio-refracting regions: $f\sim 10^{-4}$ to $5\cdot 10^{-3}$ 
(Walker \& Wardle, 
1998). From this point of view $\gamma$-rays produced via interaction of high 
energy cosmic rays (CR) with nucleons of dense clumps through the 
process $p+p\to p+p+\pi^0\to p+p+2\gamma$ are a unique probe of  
baryons hidden in dense H$_2$ clumps. 

Recently several groups have studied the $\gamma$-ray emission from 
H$_2$ clumps in detail using EGRET 
($E>100$ MeV photons) data (Chary \& Wright 1999, De Paolis et al. 
1999, Dixon et al. 1998, and Sciama 1999). 
An essential assumption of all these attempts is that 
the dark matter in the halo is optically thin to both, 
the exposing high-energy 
protons, and the resulting $\gamma$-ray photons. However, as we will 
argue below, most of the possible baryonic dark matter candidates are 
dense and 
compact enough to allow $\gamma$-ray emission only from thin external skin 
layers, and thus the existing EGRET data can trace a small fraction 
$f_{\rm T}$ of 
baryons in the halo only. In this {\it Letter} we determine  
the $\gamma$-ray background emission from small optically thick H$_2$ 
clouds. 

\section{$\gamma$-ray emission from dense clumps} 

At present, no direct observations of cold, $T \sim 3 K$, 
dense molecular clumps which might 
carry a considerable fraction of baryonic dark matter do exist, and their 
parameters, such as masses and radii, may vary in a wide range. 
Pfenniger \& Combes (1994) have argued that a considerable fraction of baryonic 
dark matter in galaxies can be accounted for by dense clumps of 
predominantly molecular hydrogen of Jovian masses and radii of 30 AU. 
Gerhard \& Silk (1996) estimated masses of 
$\sim 1~\msun$ and radii of $\sim 0.03 - 0.1$ pc. 
Walker \& Wardle (1998) have shown that the extreme scattering events might 
be naturally explained if the clumps have radii of 
$\sim 3$ AU. Larger radii, $\sim 10$ AU, are proposed by Draine (1998) 
from consideration of optical lensing of stars by dense gas clouds. Quite 
different arguments, based on the analysis of turbulent motions of HI gas in 
the halo, lead Kalberla \& Kerp (1998, KK98) to conclude  
that a significant 
fraction of the dark matter in the halo can be in form of gas clouds with 
masses of $\la 2\times 10^{-3}~\msun$ and radii of $\sim 10$ AU. 
To distinguish between these different proposals, we need to check whether 
optical depth effects might affect the observed $\gamma$-ray emission. 
High energy ($>100$ MeV) CR protons in H$_2$ are attenuated by a factor 
$e$ for $\Sigma_{\rm CR}\simeq 40$ g cm$^{-2}$ (Salati et al., 1996). The mass 
column density of a clump along the radius is 
$N_{\rm c}=4.2\times 10^6 M/r_{\rm c}^2$ g 
cm$^{-2}$, where $r_{\rm c}$ is the clump radius in AU, $M$, 
its mass in $\msun$. Using the parameters as proposed by 
Walker \& Wardle (1998) we find from this crude estimate that 
$N_{\rm c}$ exceeds $\Sigma_{\rm CR}$ by two orders of magnitude. 
$\gamma$-photons produced inside the clumps suffer from absorption. 
According to Salati et al. (1996), an optical depth of one 
in H$_2$ is reached at $\Sigma_{\rm p}\simeq 80$ g cm$^{-2}$, twice as big as 
$\Sigma_{\rm CR}$.

After this first estimate we define $f_{\rm T}$ as the fraction of 
$\gamma$-ray emission from a dense clump relative to the 
emission from an optical thin cloud. We derive $f_{\rm T}$ by integrating 
the $\gamma$-ray emission from the sphere assuming that the 
in-falling cosmic rays are distributed isotropically. This 
in turn results in an isotropic distribution for the  $\pi^o$-photons.
We assume that the density $\rho$ of the clump is 
constant and derive   
\be
f_{\rm T}={3L_{\rm CR}^2L_{\rm p}^2\over 4 r_{\rm c}^5}
\int\limits_0^{r_{\rm c}} dr F_{\rm CR}(r)F_{\rm p}(r),
\label{fraction}
\ee
where $L_i=\Sigma_i/\rho$, $i={\rm CR,~p}$, 
\ba
F_i(r)=e^{-(r_{\rm c}-r)/L_{\rm i}}[1+(r_{\rm c}-r)/L_i]
\nonumber\\ 
-e^{-(r_{\rm c}+r)/L_i}
[1+(r_{\rm c}+r)/L_i].
\ea
$f_{\rm T}$ is an approximation only, since we use frequency averaged 
attenuation lengths $\Sigma_{\rm CR}$ and $\Sigma_{\rm p}$, we neglect also  
electron Bremsstrahlung. 

In Fig. 1 we display $f_{\rm T}$ as a function of the clump radius 
$r_{\rm c}$ for
clumps with masses $10^{-4}$ to $10^{-2} \msun$. 
There is significant absorption for radii $r_{\rm c}$ $\la$ 30 AU.
Clouds described by Pfenniger \& Combes 
(1994) and by Gerhard \& Silk (1996) are transparent 
and therefore in these models the $\gamma$-ray intensity is 
proportional to the total mass contained in dense clouds and clumps. 
On the contrary, dense clumps in models described by 
Walker \& Wardle (1998), Draine (1998), and KK98 are optically thick.  
In this case
the determination of the mass of a baryonic dark matter halo from
the observed $\gamma$-ray emission depends on $f_{\rm T}$. 

\begin{figure}[htb]
\centerline{
\psfig{figure= 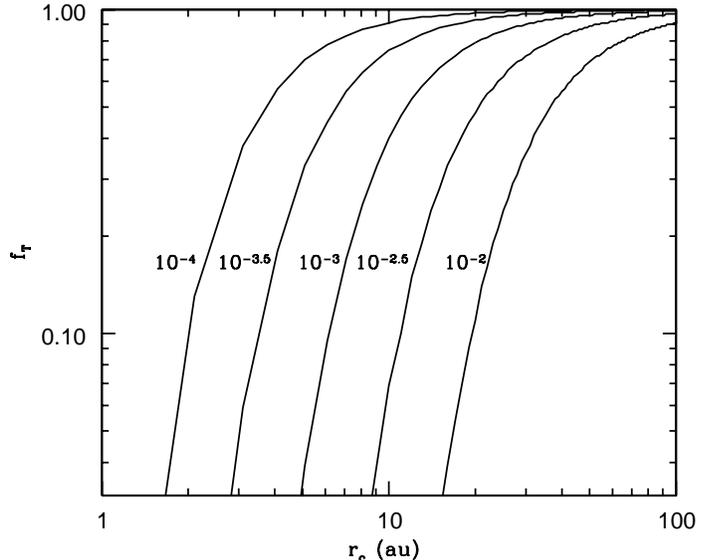,width=9.0cm,%angle= -90,
bbllx=90pt,bblly=255pt,bburx=480pt,bbury=580pt
}}

\caption{$f_{\rm T}$, the fraction of the $\gamma$-ray emission from a clump 
with radius $r_{\rm c}$ relative to the emission from an optical thin H$_2$ cloud.
The curves are for clump masses $10^{-4}$ to $ 10^{-2} \msun$. 
\label{sphere}}
\end{figure}

The $\gamma$-ray emission is caused by nuclear 
interactions between cosmic rays and matter (most prominent the 
$\pi^0$-decay), by electron Bremsstrahlung, and by inverse Compton 
interactions. We used the source functions for nucleon-nucleon 
interactions and Bremsstrahlung as published by Bertsch et al. (1993). 
For the inverse Compton interaction we have used the ``galprop'' 
database according to Strong \& Moskalenko (1997).  
The total $\gamma$-ray flux observed on Earth is a superposition of fluxes 
produced by dark matter halo clouds and diffuse components of 
the interstellar gas: the extra-planar diffuse ionized gas (Dettmar, 1992), 
the \hi disk, and the gaseous halo with \hi 
gas and plasma (KK98). The density $\rho_i(R_{\rm g},z)$,  
($R_{\rm g} = \sqrt{x^2+y^2}$), of the various gaseous components is given by 
\be
\label{distrib}
\rho_i(R_{\rm g},z)= n_i g_1(R_{\rm g})  \exp \left[ \frac{-\Phi(z)}{
\sigma_i^2\;(1+\alpha_i+\beta_i)} \right]
\ee
with $n_i$, the local midplane density of the individual component,  
$\Phi(z)$ the gravitational potential, 
$\sigma_i$ the corresponding velocity dispersion, and $\alpha_i$ and $\beta_i$
according to Parker (1966) quotients which determine the pressure of the 
magnetic field and cosmic rays relative to the gas pressure. According to 
KK98 {\em all} gaseous components have a common radial density 
distribution according to 
\begin{equation}
\label{radis}
g_1(R{\rm _g}) = $sech$^2(R_{\rm g}/A_1) / $sech$^2(R_{\odot}/A_1),
\end{equation}
with a radial scale length $A_1 = 15$ kpc. 
This relation was introduced by Taylor \& Cordes (1993) to describe 
the diffuse ionized gas component. 
The scale height of the gaseous halo according to KK98 is $h_{\rm z} = 4.4$ kpc. 

A widely used standard expression to describe the density distribution in the 
galactic halo is (e.g. De Paolis et al., 1999)
\be
\label{disdep}
\rho_{\rm H_2}(x,y,z)=\rho_0(q){a^2+R_0^2\over
a^2+x^2+y^2+(z/q)^2},
\ee
where $x,~y,~z$ are the galactocentric coordinates, $R_0$ is the Solar
distance, $\rho_0(q)$, the local dark matter density.

\section{Model calculations} 
We discuss here several models for the distribution 
of diffuse gas components and the dark matter clumps in detail: 
\begin{itemize}
\item[1)] the disk, the extra-planar diffuse ionized gas and 
the (observed) gaseous halo 
with parameters as in KK98 model according to Eq. (\ref{distrib}) and 
(\ref{radis}). 
The local midplane density of the 
$\gamma$-ray emitting gas is $n_0^{\rm h}=0.0025$ cm$^{-3}$, and the CR density 
is derived from the pressure equilibrium with the observed gas. 
In this model there is {\em no} significant $\gamma$-ray emission 
from the halo.

\item[2)] same as the previous model with the best fit local midplane density 
of the halo $\gamma$-ray gas $n_0^{\rm h}=0.065$ cm$^{-3}$. In this model the 
flat rotation curve demands a local total midplane density 
$n_0^{\rm d}=0.7$ cm$^{-3}$ (KK98). Assuming that the total dark matter 
in the Milky Way is contained in H$_2$ clumps, 
we obtain $f_{\rm T} = 0.09$. 

\item[3)] the dark matter halo with a distribution in the form 
(\ref{disdep}) with $a=5.6$ kpc, $q=0.3$, $R_0=8.5$ kpc, and a constant 
CR energy density in the halo of 0.12 eV cm$^{-3}$. This model is similar to 
the model of De Paolis et al., however without a central hole in the dark 
matter distribution at $R<10$ kpc. We fit a local 
midplane density $n_0^{\rm h}=0.55$ cm$^{-3}$. For a rotation velocity of 
$\vsun = 220$ kms$^{-1}$ we derive $n_0^{\rm d}=0.5$ cm$^{-3}$, hence 
$n_0^{\rm h} / n_0^{\rm d}= 1.1$. According to Eq. (1)  $f_{\rm T} \le 1$, 
and within the errors we obtain for this model $f_{\rm T} = 1$. 

\item[4)] same as the previous model but with $q=1$ and $n_0^{\rm h}=0.18$ 
cm$^{-3}$, we derive $n_0^{\rm d}=0.24$ cm$^{-3}$ corresponding to 
$f_{\rm T} = 0.75$.  

\item[5)] same as the model 4, but with a central hole for  
$R<10$ kpc. We determine $n_0^{\rm h}=0.35$ cm$^{-3}$. 
This is the model proposed by De Paolis et al. (1999) with 
$n_0^{\rm d}=0.32$ cm$^{-3}$. 
This model cannot reproduce the rotation velocity observed 
in the inner galaxy; $f_{\rm T}$ is undefined.
\end{itemize} 
We estimate the errors in the determination of $n_0^{\rm h}$ and 
$n_0^{\rm d}$ to 
about 10\%. The isotropic $\gamma$-ray background was fitted  
to $5 (\pm 1)\cdot 10^{-6} {\rm cts \; s^{-1} cm^{-2} sr^{-1}}$, 
in good agreement with the rate of 
$4 (\pm 1)\cdot 10^{-6} {\rm cts \; s^{-1} cm^{-2} sr^{-1}}$ due to blazars 
as determined by Mukherjee \& Chiang (1999).

\begin{figure}[htb]
\centerline{
\psfig{figure=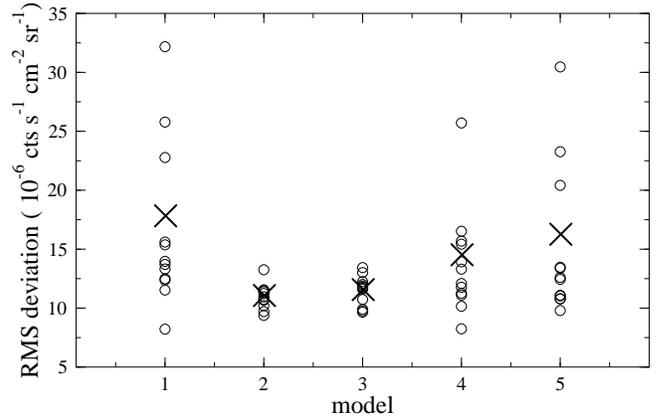,width=9.0cm,bbllx=60pt,bblly=170pt,bburx=460pt,bbury=390pt
}}
\caption{RMS deviations between observations and model calculations 
as described in Sect. 3.
The RMS scatter derived for individual scans at constant longitudes 
is displayed as circles, the total RMS deviation between model and data
is given by the crosses. 
\label{deviation}}
\end{figure}

\section{Discussion and conclusions} 
For all models we calculated the RMS deviation between data  
and model, individually for each of the scans at constant longitude 
and further for all of the data under consideration (Fig. 3). 
We found that the RMS values derived this way 
were for all of the models biased by a number of positions with significant 
deviations between model and data. Predominantly the deviations were due to 
$\gamma$-ray excess from point sources, but also from extended regions 
like the Orion complex which are not part of the model (see Fig. 3 at 
longitude $l = 180\degr$). At a few positions the emission was found to be 
systematically low. 
We decided to disregard these isolated regions and excluded  6 \% of the data, 
identical for all models, from the RMS determination.

Fig. 2 presents the derived RMS deviations for our models.
Model 1 is not a fit but gives the residual $\gamma$-ray 
emission after subtracting the $\gamma$-ray emission from disk and 
diffuse ionized gas. 
For the models 2 to 5 the additional emission 
originating from a baryonic halo has been calculated. 
Model 2 represents the best fit. The RMS deviation between data and 
model (crosses) as well as the scatter between individual scans  
at constant longitudes (circles) is minimal.
In Fig. 3 we plot model 2 in comparison with the observations.

\begin{figure}%[htbp]
\centerline{
\psfig{figure=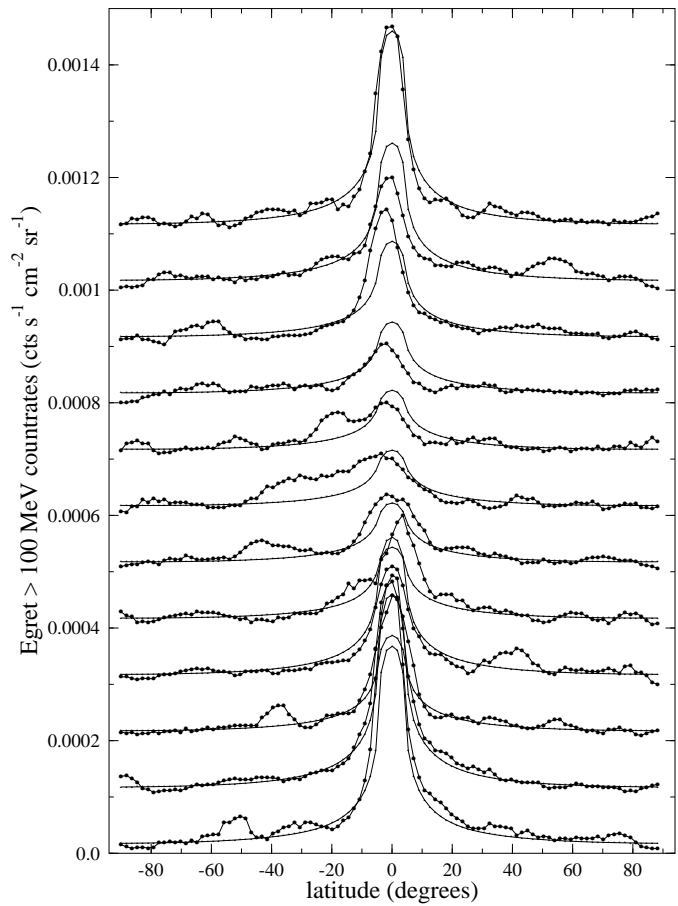,width=9.0cm,bbllx=55pt,bblly=205pt,bburx=460pt,bbury=750pt
}}
\caption{The EGRET diffuse $\gamma$-ray emission ($E_\gamma>100$ MeV) 
observations -- lines with dots, and best fit model 2 -- solid 
lines. The longitude $l$ varies from 0$^\circ$ (bottom) to 
300$^\circ$ (top) in steps of $\Delta l=30^\circ$. }
\label{gflux}
\end{figure}

The models which have been represented in Fig. 2 have been supplemented by 
additional calculations with various core radii $a$ and 
flattening parameters $q$ according to (\ref{disdep}). We found 
$0.2 \la q \la  0.4$ and $a \sim 5.6$ kpc to fit the observations well, 
however, in no case we could recover the best fit results of model 2. 
A flattening parameter $q = 0.3$ corresponds to the flattening 
of model 2, which has no free parameters concerning 
the shape of the halo. 

Halos with a flattening parameter $q \ga  0.4$ barely fit the 
observations. In particular, spherical halo models, 
as represented by model 4 and 5, result in very poor fits. 
%Our results do not support a hollow halo without dark matter in the central region for galactocentric radii $R < 10$ kpc.
%In particular, we cannot reproduce the maximum of the $\gamma$-ray emission in direction to longitude $l = 90\degr$ which was derived by De Paolis et al. (1999a,b). 
Flat dark matter models with $q \la  0.2$ or models with a scale height 
$h_{\rm z} \la 1 $ kpc for the CR distribution 
(Combes \& Pfenniger 1996) also do not fit the EGRET data in a 
satisfactory way.

Concerning the question, whether the $\gamma$-ray emission from H$_2$ 
clumps suffers from obscuration as defined by $f_{\rm T}$ in Eq. (1), we 
need to distinguish two cases, the transparent and the opaque model. 
%\begin{itemize} 
%\item 
For model 3 we derive $f_{\rm T} = 1$.
From Fig. 1 it is obvious that the radii of the clumps 
must be large, $r_{\rm c} \ga 20$ AU. Such models have been detailed by 
Pfenniger \& Combes (1994), by Combes \& Pfenniger (1996), 
and by Gerhard \& Silk (1996).
%\item  
For model 2 we derive $f_{\rm T} = 0.09$, the opaque case. 
The clumps have masses of $\sim 10^{-3} \msun$ and radii $r_{\rm c} \sim 6 $ AU.
For masses of $0.3 - 3 \cdot 10^{-3} \msun$ we derive  radii between 
3 and 10 AU respectively. The lower values are 
close to the radii estimated by Walker \& Wardle (1998) from their 
explanation of extreme scattering events. Our upper limit corresponds 
to the estimates by Draine (1998) and by KK98.
%\end{itemize}

The major difference between model 3 and 2 is the 
assumption concerning the distribution of cosmic rays on large scales. 
Models 3 to 5 are based on a constant energy density of 
0.12 eV cm$^{-3}$ out to distances of 100 kpc (De Paolis et al. 1999). 
In model 2 it is assumed that the cosmic rays are 
in pressure equilibrium with the observed gaseous halo. This results 
in a rather narrow distribution with an exponential scale height 
of $h_{\rm z} \sim$ 4.4 kpc. According to current diffusion models, 
the distribution of cosmic rays is restricted to $z$-scales of 2-4 kpc 
(Webber \& Soutoul 1998) or to $4.9^{+4}_{-2}$ kpc 
(Ptuskin \& Soutoul 1998). Similar parameters 
were used by Salati et al. (1996). Their conclusion, that only about 3 \% 
of the $\gamma$-ray flux which is expected for a gaseous dark matter halo 
can be observed is in good agreement with our determination of $f_{\rm T}$.
Strong et al. (1999) favor CR re-acceleration and 
determine $z$-scales of 4 - 10 kpc. In this case also the 
inverse Compton emission at high latitudes would be affected. 
Using parameters as proposed by Strong et al. (1999) for $z$-scales 
of 4 - 10 kpc leads to an increase of the RMS deviation between model 
and data. Clump radii, however, are affected by $\la 10$\% due to the 
steep gradient of $f_{\rm T}$ (Fig. 1).

Since there is no observational evidence for a cosmic ray halo 
at $z$-scales exceeding 10 kpc, we adopt our best fit model 2. 
We interpret the residual observed $\gamma$-ray emission 
after subtraction of the emission from disk and diffuse ionized gas layer 
as due to H$_2$ clumps with masses of $\sim 10^{-3}\msun$ 
and characteristic 
radii of $r_{\rm c} \sim 6 $ AU. Such clumps, exposed to cosmic rays, are optical 
thick and emit $\gamma$-rays only close to their surfaces. 
The Milky Way dark matter halo may contain $\sim 10^{14}$ such 
H$_2$ clumps with a total mass of $\sim 2 \cdot 10^{11} \msun$.  
The local surface column density of these clumps then is 
$\Sigma \sim 140 \; {\rm M_{\odot}\; pc^{-2}}$ (KK98).

\begin{acknowledgements} 
We thank R. Schlickeiser, J. Kerp, M. Walker, M. Pohl, A. Schr\"oer, 
and D. Pfenniger 
for helpful discussions, comments, and criticism. 
YS acknowledges the hospitality of Astronomisches Institut, 
Ruhr-Universit\"at 
Bochum (supported by DFG trough SFB 191) and Osservatorio Astrofisico
di Arcetri. 
\end{acknowledgements} 

\bs


\begin{thebibliography}{}

\bibitem{} Bertsch D.L., et al., 1993, ApJ 416, 587

\bibitem{} Chary R., Wright E. L., 1999, in The Third Stromlo Symposium: The 
Galactic Halo, eds B.K. Gibson, T.S. Axelrod, \& M.E. Putman, 
ASP Conf Ser, Vol. 165, p. 357 

\bibitem{} Combes F., Pfenniger D., 1996, in ``New Extragalactic 
Perspectives in the New South Africa", ASSL 209, eds. 
D.L, Block \& J.M. Greenberg, p. 451 

\bibitem{} De Paolis F., Ingrosso G., Jetzer Ph., Roncadelli M., 1999, 
ApJ, 510, L103

%\bibitem{} De Paolis F., Ingrosso G., Jetzer Ph., Roncadelli M., 1999b, astro-ph/9906083

\bibitem{} Dettmar R.-J., 1992, Fund. Cosm. Phys., 15, 143 

\bibitem{} Dixon D. D., Hartmann D. H., Kolaczyk E. D., et al., 1998, New Astronomy 3, 539 
%Samimi J., Diehl R., Kanbach G., Mayer-Hasselwander, Strong A. W., 1998, New Astronomy 3, 539 

\bibitem{} Draine B. T., 1998, ApJ, 509, L41 

\bibitem{} Fiedler R. L., Dennison B., Johnston K. J., Hewish A., 1987, Nature, 
326, 675 

\bibitem{} Gerhard O., Silk J., 1996, ApJ, 472, 34

%\bibitem{} Hunter S.D., et al., 1997, ApJ 481, 205

\bibitem{} Kalberla P. M. W., Kerp J., 1998, A\&A, 339, 745 (KK98)

\bibitem{} Mukherjee R., Chiang J., 1999, astro-ph/9902003 

%\bibitem{} Pfenniger D., Combes F., Martinet L., 1994, A\&A, 285, 79

\bibitem{} Pfenniger D., Combes F., 1994, A\&A, 285, 94

\bibitem{} Ptuskin V.S., Soutoul A., 1998, A\&A, 337, 859 

\bibitem{} Salati P., et al., 1996, A\&A, 313, 1

\bibitem{} Sciama, D.W., 1999, astro-ph/9906159 

\bibitem{} Strong A.W., Moskalenko I.V., 1997, 
in Proc. ${\rm4^{th}}$ Compton Symposium AIP 410, 1162

\bibitem{} Strong A.W., Moskalenko I.V., Reimer O., 1999, in Proc. 
${\rm26^{th}}$ ICRC, Vol. 4, 52

\bibitem{} Taylor J.H., Cordes J.M., 1993, ApJ 411, 674

\bibitem{} Turner M., 1999, PASP 111, 264

\bibitem{} Walker M., Wardle M., 1998, ApJ, 498, L125 

\bibitem{} Webber W.R., Soutoul A., 1998, ApJ, 506, 335 


\end{thebibliography}
\end{document}